\documentclass[
    ,final            
  ]
  {aipproc}
\usepackage{latexsym,amsmath,amssymb,wasysym}
\layoutstyle{6x9}

\begin{document}

\title{Hawking tunneling and boomerang behaviour of massive particles with $E < m$}

\classification{04.70.-s, 04.62.+v, 04.70.Dy}
\keywords      {Black Holes, Hawking Radiation, Semiclassical Approximation}

\author{Gil Jannes}{
  address={Low Temperature Laboratory, Aalto University School of Science, PO Box 15100, 00076 Aalto, Finland}
}

\author{Thomas G. Philbin}{
  address={School of Physics and Astronomy, University of St Andrews, North Haugh, St
Andrews, Fife KY16 9SS, Scotland, UK}
}

\author{Germain Rousseaux}{
  address={Universit\'e de Nice Sophia Antipolis, Laboratoire J.-A. Dieudonn\'e, UMR CNRS-UNS 6621, Parc Valrose, 06108 Nice Cedex 02, France}
}

\begin{abstract}
Massive particles are radiated from black holes through the Hawking mechanism together with the more familiar radiation of massless particles. For $E \geq m$, the emission rate is identical to the massless case. But $E < m$ particles can also tunnel across the horizon. A study of the dispersion relation and wave packet simulations show that their classical trajectory is similar to that of a boomerang. The tunneling formalism is used to calculate the probability for detecting such $E < m$ particles, for a Schwarzschild black hole of astrophysical size or in an analogue gravity experiment, as a function of the distance from the horizon and the energy of the particle.
\end{abstract}

\maketitle


\section{Introduction}

The standard vision on Hawking radiation (HR) of massive particles dates back to~\cite{Page:1976df}. It states essentially that there is a cut-off in the emission of massive particles because the Hawking temperature $T_H$ of the black hole must deliver $E \geq m$. Therefore, as anticipated in~\cite{Hawking:1974sw}, ``there will not be much emission of particles of rest mass $m$ unless the temperature $\kappa/2\pi$ is greater than $m$''. The example of an electron gives $m_e \approx 10^{-30}kg$, hence $T \sim 10^9 K$ and so the black hole mass must decrease to $M\sim 10^{-16} m_{\astrosun}\sim 10^{21} m_{Pl}$ ($m_{\astrosun}$ and $m_{Pl}$ are the solar and Planck mass, respectively) before electrons can be emitted in any significant amount. Through the emission of massless particles, the black hole will evaporate and this threshold eventually be reached. However, during most of the lifetime of an astrophysical black hole, there will be no significant emission of massive particles.

This standard vision sees HR as a ``black box'' whose outcome is measured at asymptotic infinity. It must be amended if one views HR as a near-horizon process. This allows, e.g., to reconcile the two heuristic interpretations of HR as mode conversion between positive/negative-norm  partners, related to pair creation (from vacuum fluctuations) just \emph{inside} the horizon, or just \emph{outside} the horizon. Obviously, any calculation should give the same result in both interpretations. A thermal spectrum is indeed detected at asymptotic infinity in both cases. But the mass decrease of the black hole is due to the emission of the positive partner in the first case, and to the absorption of the negative partner (in the second). There might be good reasons for the emission of positive partners to be subject to the threshold $E\geq m$, but there is no reason why the absorption of negative partners should obey a similar condition. How can these two heuristic visions then be equivalent?


\section{Dispersion relation}
The simplest dispersion relation for massive modes/particles is (in units $\hbar=1$)
\begin{equation}\label{dispersion-relation}
(\omega-Uk)^2=m^2+c^2k^2,
\end{equation}
where $U(r)<0$ is the free-fall velocity of an observer starting at rest at infinity. \mbox{$U(r)=-c\sqrt{r_h/r}$} for a Schwarzschild black hole (with $r_h$ the horizon or Schwarzschild radius, i.e. $U(r_h)=-c$), and we have defined the ``mass'' $m=\bar m c^2$. Such a Klein-Gordon dispersion relation can be derived for example from the Painlev\'e-Gullstrand-Lema\^itre form of the Schwarzschild metric
\[
ds^2=[c^2-U(r)^2]dt^2 - 2U(r) dt\, dr -dr^2 ,
\]
which provides an intuitive analogy with sound or surface waves counter-propagating with local velocity $c$ against a background fluid flowing at a speed $U$ in the laboratory frame. $\omega_0=\sqrt{m^2+k^2}$ is then the frequency in the co-moving reference frame, and $\omega=\omega_0+\bf k\cdot \bf U$ is the Doppler-shifted frequency (corresponding to the laboratory or ``black hole rest frame''). Note that $\omega$ is a conserved quantity (for stationary spacetimes), and therefore provides a good basis for a semiclassical treatment (see below).

\begin{figure}
$\begin{array}{cccc}
  \includegraphics[height=.18\textheight]{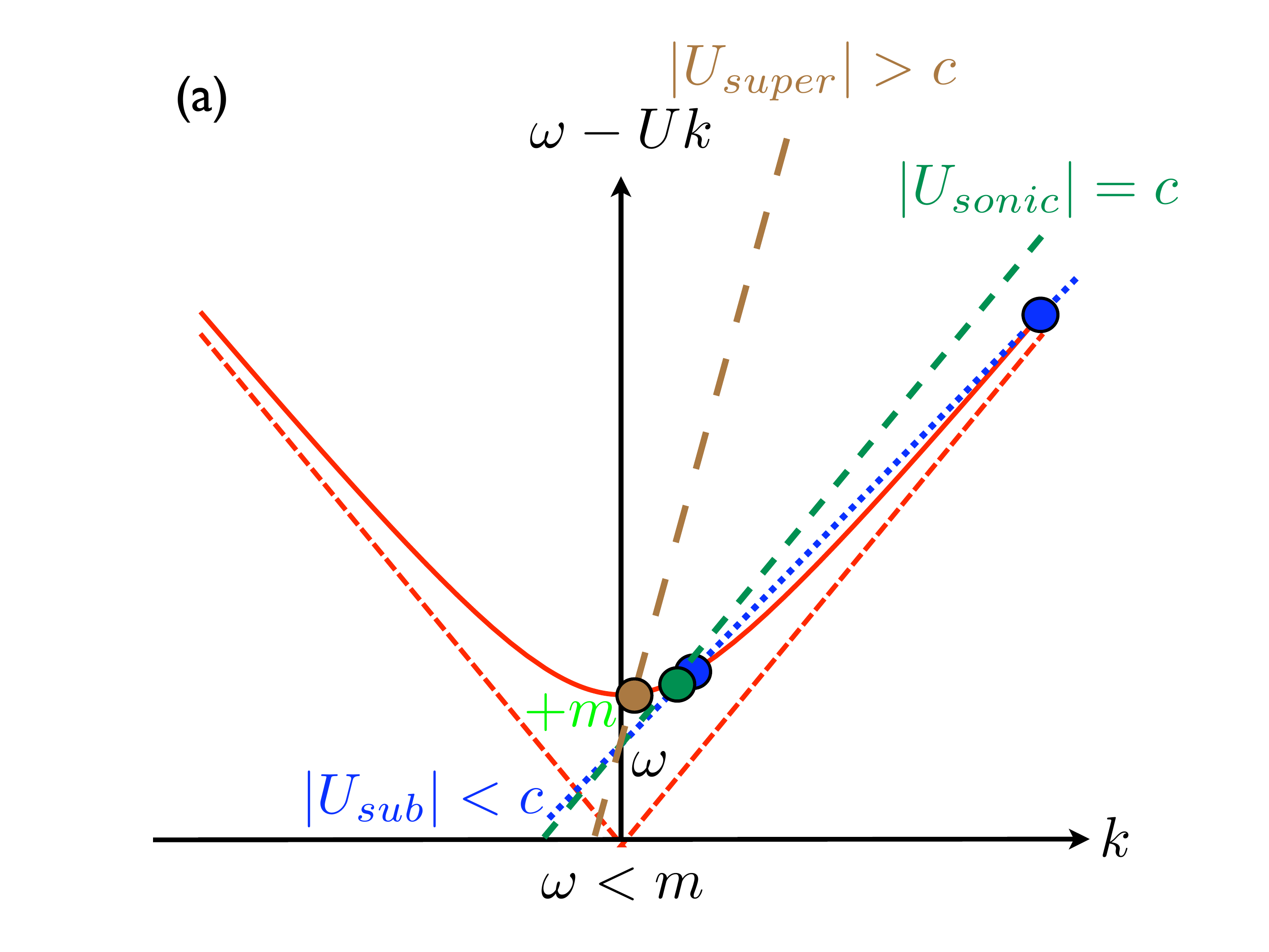}&
  \includegraphics[height=.18\textheight]{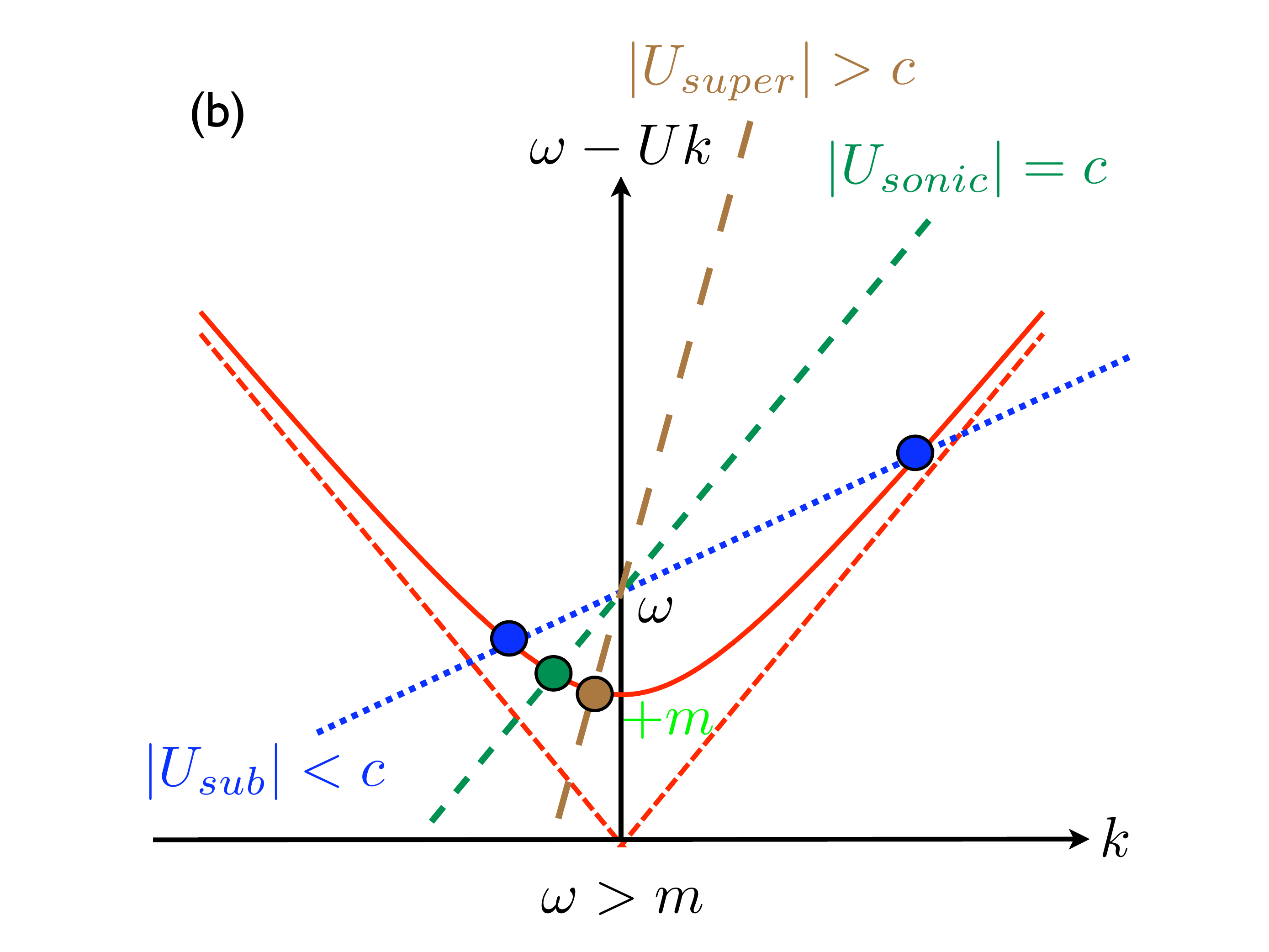}&
\\
\includegraphics[height=.18\textheight]{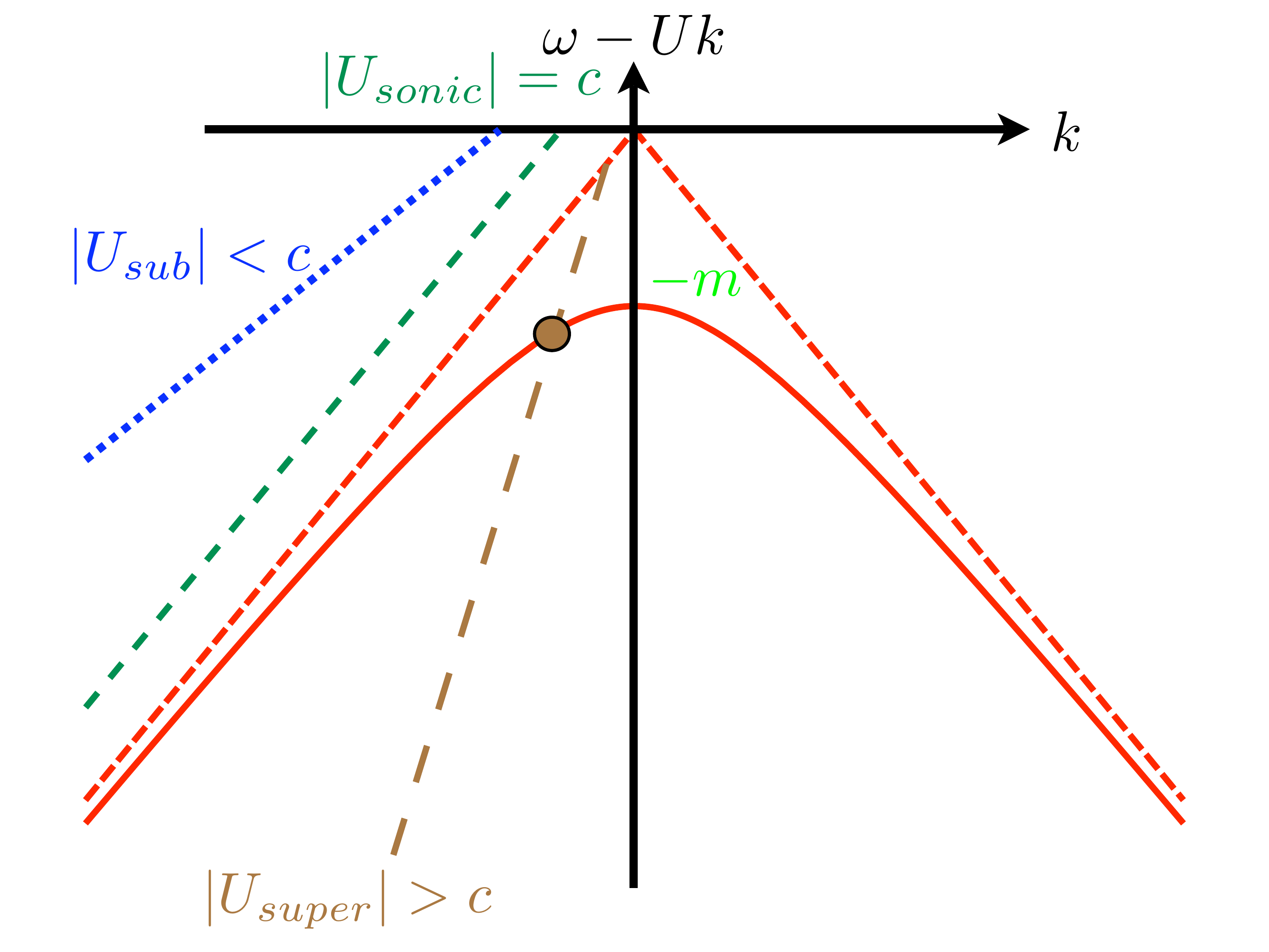}&
\includegraphics[height=.18\textheight]{dpn}
\end{array}$
\caption{Positive and negative mode solutions for the dispersion relation~\eqref{dispersion-relation} for different values of the ``counter-current velocity'' $U$ and (a) $E<m$; (b) $E>m$. The mechanism of appearance of negative-norm solutions (bottom) is independent of the ratio $E/m$, and actually equal to the massless case~\cite{Jannes:2011vb}.}\label{Fig:dispersion-relation}
\end{figure}

A graphical analysis of the mode conversion characteristic of Hawking radiation for this dispersion relation is given in Fig.~\ref{Fig:dispersion-relation}~\cite{Jannes:2011vb}. The key point here is that the appearance of the negative mode solution (negative co-moving frequency, i.e.: bottom part of the figure) is identical for the various upper parts of the figure, i.e.: the positive-negative mode conversion around the horizon is totally independent of the ratio $E/m$ (there is no threshold $E \geq m$) and in fact identical to the case of a massless mode~\cite{Jannes:2011vb}. 

It is also easy to realize that there exists a critical value $U^*$ for the counterflow in the case $E<m$, such that there are two positive mode solutions when $|U|>|U^*|$, and there is none left when $|U|<|U^*|$. $U^*$ constitutes a turning point or saddle-node bifurcation and its value is $U^*=\pm c\sqrt{1-\left(\frac{\omega}{m}\right)^2}$~\cite{Jannes:2011vb}. This shows that the $E<m$ modes cannot escape to infinity but eventually turn around like a boomerang at a location $r^*=r_h(c/U^*)^2$ (which depends on their ratio $E/m$ through $U^*$) and fall back into the black hole (see~\cite{Jannes:2011vb} for a wave packet simulation). Therefore, they do not contribute to the evaporation of the black hole. Nevertheless, they can (at least in principle) be detected at any finite distance from the horizon. Putting in numbers gives $r^*\approx 3r_h$ for $E=0.8m$, or $r^*\approx 50r_h$ for $E=0.99m$. Taking into account that, for a solar-mass black hole, $r_h\sim 3km$, and moreover that the resolution of any detection will be limited by the time until re-absorption in the black hole, it seems that this boomerang effect only has a purely theoretical interest for astrophysical black holes (except perhaps for supermassive ones).

\section{Tunneling formalism}
The tunneling probability $W$ in a semiclassical formalism is $W\propto \exp[-2\text{Im}S]$, where $\text{Im}S=\text{Im}\int p_r(r)dr $ is the imaginary part of the action along the semiclassical trajectory, and $p_r(r)$ is obtained from the dispersion relation~\eqref{dispersion-relation}, which we now write \mbox{$(E-{\bf p}\cdot {\bf U})^2=m^2+p^2$}. The expected final result is of the form $W(E)\propto \exp[-2\text{Im}S_1]\exp[-2\text{Im}S_2] $, where $S_1$ corresponds to tunneling through the horizon, and $S_2$ stems from the second classically prohibited region beyond the horizon in the case $E<m$. Recall that for $U(r \to \infty)=0$, $E^2=m^2+p^2$ (i.e., the standard Minkowski dispersion relation), and hence particles with $E<m$ are certainly classically forbidden at flat asymptotic infinity. From the dispersion relation, $p$ can be written 
\[
p
=-\frac{EU}{1-U^2} + \frac{1}{1-U^2}\sqrt{m^2(U^2-1)+E^2}\\
=p_1+p_2.
\]
$S_1$ is obtained from $p_1$ in the usual way~\cite{Volovik:1999fc}. We shift the contour of integration to the complex plane, apply the residue theorem $\int_C f(z)=2\pi i\text{Res} f$, find a pole along the radial path: $U=-1$ at $r=r_h$, and obtain $\text{Res} p_1=\text{Res}\frac{-EU}{(1+U)(1-U)}=\frac{E}{2U'(r_h)}$. This leads to the standard Hawking result: $\tilde{W_1}(E)\propto \exp[-E/T_H]$, with $T_H=|U'(r_h)|/2\pi$, where the prime denotes $d/dr$. This confirms that the presence of a mass has absolutely no influence on the probability of tunneling across the horizon.

\begin{figure}
  \includegraphics[height=.16\textheight]{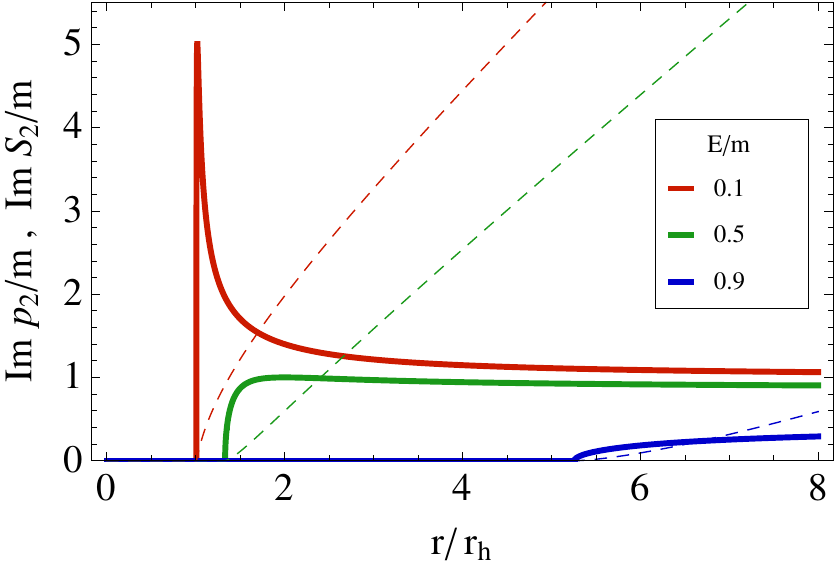}
  \caption{Semiclassical barrier Im$p_2$ (thick lines) and tunneling action $\text{Im}S_2$ (dashed lines) beyond the horizon $r_h$ for various values of $E/m<1$~\cite{Jannes:2011qp}.}\label{Fig:S2}
\end{figure}

We now consider a detector at a distance $R$ and examine the second contribution $S_2$. There will be an imaginary contribution if $E^2<m^2(1-U^2)$ (i.e., $E<E_c(R)=m\left(1-\frac{r_h}{R} \right)^{1/2}$ for given $R$, or $R>r_c(E)=\frac{1}{1-\frac{E^2}{m^2}}r_h$ for given $E$). Ffor $R\to \infty$, we recover the condition $E<m$. 
For a Schwarzschild profile, we obtain $\text{Im}S_2 =\int_{r_c}^R dr \frac{1}{1-r_h/r}\sqrt{m^2(1-r_h/r)-E^2}$, see Fig.~\ref{Fig:S2}. In the limit $R\gg r_h$, the integral is dominated by the contributions $U\to 0$ and one can write 
$\text{Im}S_2=\sqrt{m^2-E^2}\int_{r_c}^R dr\approx R \sqrt{m^2-E^2}$ (assuming $R\gg r_c$). The overall tunneling rate then becomes 
\[
W(E)\propto \exp\left[\frac{-E}{T_H}\right]\exp\left[-2R\sqrt{m^2-E^2}\right],
\]
Other simple analytical results can be obtained in the near-horizon limit $R \to r_h$~\cite{Jannes:2011qp}.

\section{Analogue gravity}
As mentioned above, the emission of $E<m$ particles has little practical impact for astrophysical black holes. However, in the case of analogue-gravity systems with a massive dispersion relation, they become more interesting. Indeed, in such systems, particles can be detected inside the black hole also, so there is no limit on the spectral resolution. Moreover, the region where $U \neq 0$ can be extended far away from the horizon. Candidate systems include acoustic waves in ion rings, massive phonons from 2-component BECs, Langmuir waves in a moving plasma, barotropic waves (inertia-gravity or Poincar\'{e} waves), and spin waves in magnetic media~\cite{Jannes:2011vb}. Note that the existence of two imaginary contributions for $E<m$ indicates the presence of a double barrier, and hence the possibility of creating resonant states according to the Bohr-Sommerfeld quantization condition. These resonant states could be interesting candidates for detection and hence for a confirmation of some of the curious features of Hawking radiation described here.



\bibliographystyle{aipproc}   


\end{document}